\documentclass[a4paper, 10 pt, conference]{ieeeconf}
\IEEEoverridecommandlockouts
\usepackage{amsfonts,amsmath,amssymb} 

\newcommand{\be}{\begin{equation}}
 \newcommand{\ee}{\end{equation}}
\newcommand{\bea}{\begin{eqnarray}}
 \newcommand{\eea}{\end{eqnarray}}
 \newcommand{\beann}{\begin{eqnarray*}}
 \newcommand{\eeann}{\end{eqnarray*}}
\newcommand{\ba}{\begin{array}}
 \newcommand{\ea}{\end{array}}

\usepackage{psfrag,color}
\usepackage{enumerate,cite,latexsym,graphicx}

\newtheorem{definition}{Definition}

\title{\LARGE \bf Coherent-Classical Estimation for Quantum Linear Systems
}

\author{Ian R.~Petersen %
\thanks{This work was supported by the
Australian Research Council (ARC) and the Air Force Office of Scientific
Research (AFOSR). This material is based on research sponsored by the
Air Force Research Laboratory, under agreement numbers
FA2386-09-1-4089 and FA2386-12-1-4075.  The U.S. Government is authorized to reproduce and
distribute reprints for Governmental purposes notwithstanding any
copyright notation thereon.
The views and conclusions contained herein are those of the authors
and should not be interpreted as necessarily representing the official
policies or endorsements, either expressed or implied, of the Air
Force Research Laboratory or the U.S. Government. }%
\thanks{Ian R. Petersen is with the School of  Engineering and Information Technology, 
        University of New South Wales at the Australian Defence Force Academy, Canberra ACT 2600, Australia.
         {\tt\small i.r.petersen@gmail.com} } 
}%

\begin{document}

\maketitle
\thispagestyle{empty}
\pagestyle{empty}

\begin{abstract}
This paper introduces a problem of coherent-classical estimation for a class of linear quantum systems. In this problem, the estimator is a mixed quantum-classical system which produces a classical estimate of a system variable. The coherent-classical estimator may also involve coherent feedback. An example involving optical squeezers is given to illustrate the efficacy of this idea. 
\end{abstract}

\section{Introduction} \label{sec:intro}
In recent years, a number of papers have considered the feedback
control of systems whose dynamics are governed by the laws of quantum 
mechanics instead of classical mechanics; see e.g., \cite{YK03A,YK03B,YAM06,JNP1,NJP1,GGY08,MaP3,MaP4,YNJP1,GJ09, GJN10,WM10,PET10Ba}. 
Quantum linear systems are an important class of  quantum systems; e.g.,  see \cite{WM94}, \cite{WM10}, \cite{GZ00}, \cite{EB05}, \cite{BE08}, \cite{WD05,JNP1,NJP1,NJD09,YAM06,MAB08,YNJP1,GGY08,SSM08,GJN10}).
These linear stochastic models describe quantum optical devices such as optical cavities \cite{BR04}, \cite{WM94}, linear quantum amplifiers \cite{GZ00}, and finite bandwidth squeezers \cite{GZ00}. 

Some recent papers on the feedback control of linear quantum systems 
have considered the case in which the feedback controller  itself is
also a
quantum system. Such feedback control is often
referred to as coherent quantum control; e.g., see
\cite{WM94a,SL00,JNP1,NJP1,MaP1a,MaP3,MaP4,MAB08,GW09}. In this paper, we consider a related coherent-classical estimation problem in which the estimator consists of a classical part, which produces the final required estimate and a quantum part, which may also involve coherent feedback. A related but different problem is the problem of constructing a quantum observer; see, \cite{MJ12a}. A quantum observer is a purely quantum system which aims to produce a quantum  estimate of a variable for a given quantum plant. In contrast, we consider a coherent-classical estimator which is a mixed quantum classical system, which produces a classical estimate of a variable for a given quantum plant.  We formulate the problem of optimal coherent-classical estimation and then present an example involving optical cavities and dynamic squeezers to show that a coherent-classical estimator may yield improved performance which compared with a classical-only estimator.  

\section{Linear  Quantum Systems and Physical Realizability} \label{sec:systems}
We consider a class of linear  quantum  systems described by the quantum stochastic differential equations (QSDEs), (e.g., see
\cite{GJN10,PET10Ba,ShP5}):
\begin{eqnarray}
\label{sys} \left[\begin{array}{l} d a(t)\\d
a(t)^\#\end{array}\right] &=&  F\left[\begin{array}{l} 
a(t)\\ a(t)^\#\end{array}\right]dt +  G
\left[\begin{array}{l} d\mathcal{A}(t)
\\ d\mathcal{A}(t)^{\#} \end{array}\right];  \nonumber \\
\left[\begin{array}{l} d\mathcal{A}^{out}(t)
\\ d\mathcal{A}^{out}(t)^{\#} \end{array}\right] &=&
 H \left[\begin{array}{l}  a(t)\\
a(t)^\#\end{array}\right]dt +  K \left[\begin{array}{l} d\mathcal{A}(t)
\\ d\mathcal{A}(t)^{\#} \end{array}\right],\nonumber \\
\end{eqnarray}
where
\begin{eqnarray}
\label{FGHKform}  F =\Delta( F_1,  F_2),
 &&
 G = \Delta( G_1,  G_2), \nonumber \\
 H = \Delta( H_1, H_2),
 &&  K = \Delta( K_1, K_2).
\end{eqnarray}
Here, $ a(t) = \left[ {a_1 (t) 
\cdots a_n (t)} \right]^T$ is a vector of  annihilation operators. The adjoint of the operator $a_i$ is
  denoted by $a_i^*$
and is referred to as a \emph{creation operator}.
Also,  the notation $\Delta(F_1,F_2)$ denotes the matrix $\left[ \ba{cc}
 F_1 & F_2 \\ F_2^\# & F_1^\#
     \ea  \right]$. Furthermore, $^\dagger$ denotes the adjoint transpose of a vector of operators or the complex conjugate transpose of a complex matrix. In addition, $^\#$ denotes the adjoint of a vector of operators or the complex conjugate of a complex matrix. 
Moreover, $F_1 \in \mathbb{C}^{n \times n}$, $F_2 \in \mathbb{C}^{n \times n}$, 
$G_1 \in \mathbb{C}^{n\times m}$, $G_2 \in \mathbb{C}^{n\times m}$, 
$H_1 \in \mathbb{C}^{m \times n}$, $H_2 \in \mathbb{C}^{m \times n}$,
$K_1 \in \mathbb{C}^{m \times m}$ and $K_2 \in \mathbb{C}^{m \times m}$. 

The vector $
 \mathcal{A}=\left[\begin{array}{cccc}
\mathcal{A}_1& \mathcal{A}_2& \ldots & \mathcal{A}_m
\end{array}\right]^T
$
represents a collection of external independent quantum fields modelled by
bosonic annihilation field operators $\mathcal{A}_1,
\mathcal{A}_2,\ldots,\mathcal{A}_m$. Also, the vector $\mathcal{A}^{out}$ represents the corresponding vector of output field operators. 
For each annihilation field operator
 $\mathcal{A}_k$, there is a
corresponding creation field operator
 $\mathcal{A}_k^*$, which is
 the operator adjoint of
$\mathcal{A}_k$ (see \cite{BHJ07},
\cite{PAR92} and \cite{HP84}).
 More
details concerning this class of quantum systems can be found in the
references \cite{JNP1,GJN10,PET10Ba,ShP5}.

In the coherent classical filtering problem to be considered in this paper, we require part of the estimator to be a quantum system. In order to achieve this, we will restrict attention to quantum systems described by QSDEs of the form (\ref{sys}), (\ref{FGHKform}) which are physically realizable according to the following definition. 

\begin{definition} (See \cite{JNP1,ShP5,PET10Ba}.) 
\label{phys_real}
A complex linear quantum 
  system of the form (\ref{sys}), (\ref{FGHKform}) is 
  said to be {\em physically realizable} if there  exists a 
complex commutation matrix $\Theta= \Theta^\dagger$, a complex
Hamiltonian matrix $M = 
M^\dagger$, and a  coupling matrix $N$ such that 
\begin{equation}
\label{Psiform}
\Theta = TJT^\dagger
\end{equation}
where $T=\Delta(T_1,T_2)$ is non-singular, $M$ and $N$ are of the form
\begin{equation}
\label{tildeMN}
 M= \Delta( M_1, M_2), N= \Delta( N_1, N_2)
\end{equation}
and 
\begin{eqnarray}
\label{generalizedFGHK1}
 F &=& -\imath \Theta   M -\frac{1}{2} \Theta  N^\dagger J  N; \nonumber \\
 G &=& -\Theta   N^\dagger J; \nonumber \\
 H &=&  N; \nonumber \\
 K &=& I.
\end{eqnarray}
\end{definition} 
In this
definition, if the system (\ref{sys}) is physically realizable, then
the matrices $M$ and $N$ define a complex open harmonic
oscillator with  coupling operator
\[
L = \left[\begin{array}{cc}\tilde N_1 & \tilde N_2 \end{array}\right]
\left[\begin{array}{c} a \\  a^\#\end{array}\right]
\]
and
a Hamiltonian operator 
\[
H =
 \frac{1}{2}\left[\begin{array}{cc} a^\dagger &
       a^T\end{array}\right] M
\left[\begin{array}{c} a \\  a^\#\end{array}\right];
\]
e.g., see 
\cite{GZ00}, \cite{PAR92}, \cite{BHJ07},
\cite{JNP1} and \cite{EB05}. State space and frequency domain conditions for physical realizability can be found in \cite{JNP1,ShP5}. 

\section{Coherent-Classical Estimation}
\label{sec:estimation}
\begin{figure}[htbp]
 \begin{center}
\psfrag{MA}{$\mathcal{A}$}
\psfrag{MU}{$\mathcal{U}$}
\psfrag{z0}{$z$}
\psfrag{MY}{$\mathcal{Y}$}
\psfrag{y0}{$y$}
\psfrag{zh}{$\hat z$}
\includegraphics[width=8.5cm]{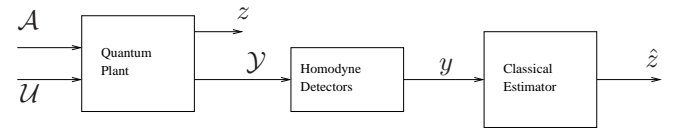}
 \end{center}
 \caption{Quantum estimation using a purely classical estimator.}
 \label{FA}
\end{figure}

In this section, we introduce a problem of coherent-classical estimation. In this problem, we begin with a quantum ``plant'' which is a quantum system of the form (\ref{sys}), (\ref{FGHKform}) defined as follows:
\begin{eqnarray}
\label{plant} 
\left[\begin{array}{l} d a(t)\\d
a(t)^\#\end{array}\right] &=&  F\left[\begin{array}{l} 
a(t)\\ a(t)^\#\end{array}\right]dt +  G_1
\left[\begin{array}{l} d\mathcal{A}(t)
\\ d\mathcal{A}(t)^{\#} \end{array}\right]  \nonumber \\
&&+  G_2
\left[\begin{array}{l} d\mathcal{U}(t)
\\ d\mathcal{U}(t)^{\#} \end{array}\right];  \nonumber \\
\left[\begin{array}{l} d\mathcal{Y}(t)
\\ d\mathcal{Y}(t)^{\#} \end{array}\right] &=&
 H \left[\begin{array}{l}  a(t)\\
a(t)^\#\end{array}\right]dt +  K \left[\begin{array}{l} d\mathcal{A}(t)
\\ d\mathcal{A}(t)^{\#} \end{array}\right];\nonumber \\
z &=& C\left[\begin{array}{l} 
a(t)\\ a(t)^\#\end{array}\right].
\end{eqnarray}
Here, $z$ denotes a scalar operator on the underlying Hilbert space which represents the quantity to be estimated. Also, $\mathcal{Y}(t)$ represents the vector of output fields of the plant which will be used by the estimator to obtain an estimate of $z$ and $\mathcal{U}(t)$ represents the control input to the plant. As above, $\mathcal{A}(t)$ represents a vector of quantum noises acting on the plant. In the case of a purely classical estimator, the control input is also taken as a vector of quantum noises acting on the plant. Also, in the case of a purely classical estimator, a quadrature of each component of the vector $\mathcal{Y}(t)$ is measured using homodyne detection to produce a corresponding classical signal $y_i$; e.g., see \cite{BR04}. This is represented by the following equations:
\begin{eqnarray}
\label{homodyne1}
dy_1 &=& \cos(\theta_1)d\mathcal{Y}_1+\sin(\theta_1)\mathcal{Y}_1^*;\nonumber \\
&\vdots& \nonumber \\
dy_m &=& \cos(\theta_m)d\mathcal{Y}_m+\sin(\theta_m)\mathcal{Y}_m^*.
\end{eqnarray}
Here, the angles $\theta_1, \ldots, \theta_m$, determine the quadrature measured by each homodyne detector. The vector of classical signals $y = \left[\begin{array}{cccc}y_1 & y_2 & \ldots & y_m\end{array}\right]^T$ is then used as the input to a classical estimator defined as follows:
\begin{eqnarray}
\label{classical_estimator1} 
 dx_e &=&  F_ex_edt +  G_edy;\nonumber \\
\hat z &=& H_e x_e.
\end{eqnarray}
Here $\hat z$ is a scalar classical estimate of the quantity $z$. Corresponding to this estimate is the estimation error
\begin{equation}
\label{error}
e = z - \hat z.
\end{equation}
Here, $e$ is an operator on the underlying Hilbert space and the second term in the expression for $e$ in (\ref{error}) is interpreted as the complex number $\hat z$ multiplied by the identity operator on the underlying Hilbert space. Then, the optimal classical estimator is defined as the system (\ref{classical_estimator1}) which minimizes the quantity
\begin{equation}
\label{cost}
J = \lim_{t \rightarrow \infty} <e^*(t)e(t)>
\end{equation}
where $< \cdot >$ denotes the quantum expectation over the joint classical quantum system defined by (\ref{plant}), (\ref{homodyne1}), (\ref{classical_estimator1}). This problem is illustrated in Figure \ref{FA}. It is straightforward to verify using a similar approach to that given in \cite{SPJ1a,NJP1} that the optimal classical estimator is given by the standard (complex) Kalman filter defined for the system (\ref{plant}), (\ref{homodyne1}). 

We now extend this problem to case of a coherent-classical estimator. In the case of coherent-classical estimation, we do not feed the plant output $\mathcal{Y}(t)$ directly into a bank of homodyne detectors as in (\ref{homodyne1}) but rather, we feed this output into another quantum system referred to as a coherent controller, which also provides coherent feedback control to the quantum plant. This coherent controller is defined as follows:
\begin{eqnarray}
\label{coherent_estimator} 
\left[\begin{array}{l} d a_c(t)\\d
a_c(t)^\#\end{array}\right] &=&  F_c\left[\begin{array}{l} 
a_c(t)\\ a_c(t)^\#\end{array}\right]dt +  G_{c1} \left[\begin{array}{l} d\mathcal{\tilde A}(t)
\\ d\mathcal{\tilde A}(t)^{\#} \end{array}\right];\nonumber \\
&&+  G_{c2}
\left[\begin{array}{l} d\mathcal{Y}(t)
\\ d\mathcal{Y}(t)^{\#} \end{array}\right];  \nonumber \\
\left[\begin{array}{l} d\mathcal{\tilde Y}(t)
\\ d\mathcal{\tilde Y}(t)^{\#} \end{array}\right] &=&
 \tilde H_c \left[\begin{array}{l}  a_c(t)\\
a_c(t)^\#\end{array}\right]dt +  \tilde K_{c1} \left[\begin{array}{l} d\mathcal{\tilde A}(t)
\\ d\mathcal{\tilde A}(t)^{\#} \end{array}\right]\nonumber \\
&&+  \tilde K_{c2}
\left[\begin{array}{l} d\mathcal{Y}(t)
\\ d\mathcal{Y}(t)^{\#} \end{array}\right]; \nonumber \\
\left[\begin{array}{l} d\mathcal{U}(t)
\\ d\mathcal{U}(t)^{\#} \end{array}\right] &=&
  H_c \left[\begin{array}{l}  a_c(t)\\
a_c(t)^\#\end{array}\right]dt +   K_{c1} \left[\begin{array}{l} d\mathcal{\tilde A}(t)
\\ d\mathcal{\tilde A}(t)^{\#} \end{array}\right]\nonumber \\
&&+   K_{c2}
\left[\begin{array}{l} d\mathcal{Y}(t)
\\ d\mathcal{Y}(t)^{\#} \end{array}\right]. \nonumber \\
\end{eqnarray}
Here, $\mathcal{\tilde A}$ represents an additional quantum noise acting on the quantum part of the coherent-classical estimator and $\mathcal{\tilde Y}$ represents its estimation output field. Note that the dimension of the estimation output field vector $\mathcal{\tilde Y}$ may be different from the dimension of the field vector $\mathcal{Y}$. The quantum system (\ref{coherent_estimator}) is required to be physically realizable. Note, that in order to meet the definition of physical realizability in Definition \ref{phys_real}, it may be necessary to augment this system with some additional unused output fields; see also \cite{MaP4,PET13Ca}. 

\begin{figure}[htbp]
 \begin{center}
\psfrag{MA0}{$\mathcal{A}$}
\psfrag{MU0}{$\mathcal{U}$}
\psfrag{MAT}{$\mathcal{\tilde A}$}
\psfrag{z0}{$z$}
\psfrag{MY0}{$\mathcal{Y}$}
\psfrag{MYT}{$\mathcal{\tilde Y}$}
\psfrag{y0}{$\tilde y$}
\psfrag{zh}{$\hat z$}
\includegraphics[width=8.5cm]{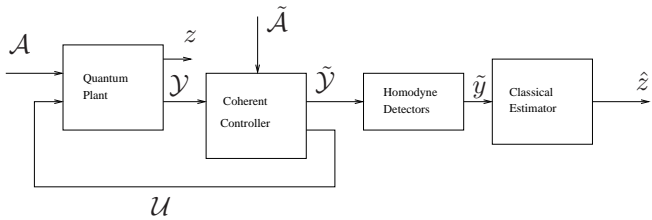}
 \end{center}
 \caption{Quantum estimation using a coherent-classical estimator.}
 \label{FB}
\end{figure}

A quadrature of each component of the vector $\mathcal{\tilde Y}(t)$ is measured using homodyne detection to produce a corresponding classical signal $\tilde y_i$; e.g., see \cite{BR04}. This is represented by the following equations:
\begin{eqnarray}
\label{homodyne2}
d\tilde y_1 &=& \cos(\tilde \theta_1)d\mathcal{\tilde Y}_1+\sin(\tilde \theta_1)\mathcal{\tilde Y}_1^*;\nonumber \\
&\vdots& \nonumber \\
d\tilde y_{\tilde m} &=& \cos(\tilde \theta_{\tilde m})d\mathcal{\tilde Y}_{\tilde m}
+\sin(\tilde \theta_{\tilde m})\mathcal{\tilde Y}_{\tilde m}^*.
\end{eqnarray}
Here, the angles $\tilde \theta_1, \ldots, \tilde \theta_{\tilde m}$, determine the quadrature measured by each homodyne detector. The vector of classical signals $\tilde y = \left[\begin{array}{cccc}\tilde y_1 & \tilde y_2 & \ldots & \tilde y_{\tilde m}\end{array}\right]^T$ is then used as the input to a classical estimator defined as follows:
\begin{eqnarray}
\label{classical_estimator2} 
 d\tilde x_e &=&  \tilde F_e\tilde x_edt +  \tilde G_ed\tilde y;\nonumber \\
\hat z &=& \tilde H_e \tilde x_e.
\end{eqnarray}
Here $\hat z$ is a scalar classical estimate of the quantity $z$.
Corresponding to this estimate is the estimation error (\ref{error}).
 Then, the optimal coherent-classical estimator is defined as the systems (\ref{coherent_estimator}),  (\ref{classical_estimator2}) which together minimize the quantity (\ref{cost}).  This problem is illustrated in Figure \ref{FB}. Note that the coherent controller is not required to directly produce an estimate of the variables of the quantum plant as in the quantum observer considered in \cite{MJ12a}. Rather, the coherent controller works only in combination with the classical estimator to produce a classical estimate of the quantity $z$. 

We can now combine the quantum plant (\ref{plant}) and the coherent controller (\ref{coherent_estimator}) to yield a closed loop  quantum linear system defined by the following QSDEs:
\begin{eqnarray*}
\left[\begin{array}{l} d a\\d
a^\#\\ d a_c\\d
a_c^\#\end{array}\right] &=&  \left[\begin{array}{cc} F+G_2K_{c2}H & G_2H_c \\ G_{c2}H & F_c \end{array}\right]\left[\begin{array}{l} 
a\\ a^\#\\a_c\\ a_c^\#\end{array}\right]dt 
\end{eqnarray*}
\begin{eqnarray*}
+  \left[\begin{array}{cc} G_1+G_2K_{c2}K & G_2K_{c1} \\ G_{c2}K & G_{c1} \end{array}\right]
\left[\begin{array}{l} d\mathcal{A}
\\ d\mathcal{A}^{\#}\\d\mathcal{\tilde A}
\\ d\mathcal{\tilde A}^{\#} \end{array}\right];
\end{eqnarray*}
\begin{eqnarray}
\label{augmented_system} 
\left[\begin{array}{l} d\mathcal{\tilde Y}
\\ d\mathcal{\tilde Y}^{\#} \end{array}\right] &=&
\left[\begin{array}{cc}\tilde K_{c2}H & \tilde H_c \end{array}\right] \left[\begin{array}{l}  a\\
a^\#\\a_c\\ a_c^\#\end{array}\right]dt \nonumber \\
&&+  \left[\begin{array}{cc}\tilde K_{c2}K & \tilde K_{c1} \end{array}\right]\left[\begin{array}{l} d\mathcal{A}
\\ d\mathcal{A}^{\#} \\d\mathcal{\tilde A}
\\ d\mathcal{\tilde A}^{\#}\end{array}\right].\nonumber \\
\end{eqnarray}

Once the coherent controller (\ref{coherent_estimator}) has been determined, it is straightforward to verify using a similar approach to that given in \cite{SPJ1a,NJP1} that the optimal classical estimator (\ref{classical_estimator2}) is given by the standard (complex) Kalman filter defined for the system (\ref{augmented_system}), (\ref{homodyne2}). Indeed, this optimal classical estimator is obtained from the stabilizing solution to the algebraic Riccati equation
\begin{eqnarray}
\label{are1}
\lefteqn{F_aP+PF_a^\dagger+G_aG_a^\dagger} \nonumber \\
&& - (G_aK_a^\dagger+PH_a^\dagger)L^\dagger(LK_aK_a^\dagger L^\dagger)^{-1}(G_aK_a^\dagger+PH_a^\dagger)^\dagger\nonumber \\
& =& 0
\end{eqnarray}
where
\begin{eqnarray}
\label{augmented_matrices}
F_a &=& \left[\begin{array}{cc} F+G_2K_{c2}H & G_2H_c \\ G_{c2}H & F_c \end{array}\right];\nonumber \\
G_a &=& \left[\begin{array}{cc} G_1+G_2K_{c2}K & G_2K_{c1} \\ G_{c2}K & G_{c1} \end{array}\right];\nonumber \\
H_a&=& \left[\begin{array}{cc}\tilde K_{c2}H & \tilde H_c \end{array}\right];~~K_a= \left[\begin{array}{cc}\tilde K_{c2}K & \tilde K_{c1} \end{array}\right];\nonumber \\
L &=& \left[\begin{array}{cc}L_1 & L_2\end{array}\right];\nonumber \\
L_1&=&\left[\begin{array}{cccc}\cos(\tilde \theta_1) & 0 & \ldots & 0\\
0 & \cos(\tilde \theta_2) & \ldots & 0\\
& & \ddots & \\
& & & \cos(\tilde \theta_{\tilde m})\end{array}\right];\nonumber \\
L_2&=&\left[\begin{array}{cccc}\sin(\tilde \theta_1) & 0 & \ldots & 0\\
0 & \sin(\tilde \theta_2) & \ldots & 0\\
& & \ddots & \\
& & & \sin(\tilde \theta_{\tilde m})\end{array}\right];
\end{eqnarray}
e.g., see \cite{KS72}. Here we assume that the quantum noises $\mathcal{A}$ and $\mathcal{\tilde A}$ are independent and purely canonical; i.e.,  $d\mathcal{A}d\mathcal{A}^\dagger = I dt$ and $d\mathcal{\tilde A}d\mathcal{\tilde A}^\dagger = I dt$ (e.g., see \cite{MaP4}). Then, the corresponding optimal classical estimator (\ref{classical_estimator2}) is defined by the equations:
\begin{eqnarray}
\label{classical_estimator_matrices}
\tilde F_e &=& F_a - G_e LH_a;\nonumber \\
\tilde G_e &=& (G_aK_a^\dagger+PH_a^\dagger)L^\dagger(LK_aK_a^\dagger L^\dagger)^{-1};\nonumber \\
\tilde H_e &=& \left[\begin{array}{cc} C & 0 \end{array}\right];
\end{eqnarray}
e.g., see \cite{KS72}. The corresponding value of the cost (\ref{cost}) is given by 
\begin{equation}
\label{cost1}
J = \left[\begin{array}{cc} C & 0 \end{array}\right]P\left[\begin{array}{c} C^\dagger\\ 0 \end{array}\right]
\end{equation}
where $P$ is the  stabilizing solution to the algebraic Riccati equation (\ref{are1}). Thus, the optimal coherent-classical estimation problem can be solved by first choosing the coherent controller (\ref{coherent_estimator})  to minimize the cost (\ref{cost1}). Then, the classical estimator (\ref{classical_estimator2}) is constructed according to the equations (\ref{classical_estimator_matrices}). 

\begin{figure}[htbp]
 \begin{center}
\psfrag{MA0}{$\mathcal{A}$}
\psfrag{MU0}{$\mathcal{U}$}
\psfrag{MAT}{$\mathcal{\tilde A}$}
\psfrag{z0}{$z$}
\psfrag{MY0}{$\mathcal{Y}$}
\psfrag{MYT1}{$\mathcal{\tilde Y}_1$}
\psfrag{MYT2}{$\mathcal{\tilde Y}_2$}
\psfrag{y0}{$\tilde y$}
\psfrag{zh}{$\hat z$}
\includegraphics[width=8.5cm]{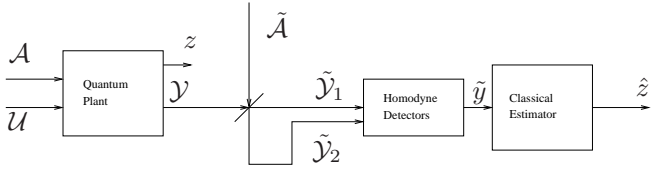}
 \end{center}
 \caption{Coherent-classical estimation using a beam splitter.}
 \label{FC}
\end{figure}

A simple example of a coherent-classical estimator arises in the case in which the plant output $\mathcal{Y}$ is a scalar operator and the coherent controller is simply a beam splitter and no feedback is used as shown in Figure \ref{FC}.
This approach is commonly referred to as dual homodyne measurement. Furthermore, it is related to an equivalent method referred to as heterodyne measurement; e.g., see \cite{BR04}. Hence, this approach is also referred to as heterodyne measurement. However, it is well known that this approach does not lead to any advantages over the purely classical estimation approach described above. In the next section, we consider the case in which the coherent controller is a dynamic squeezer and feedback is used.  We show that this does lead to advantages over the purely classical estimation approach.

\section{Dynamic Squeezer Systems} \label{sec:squeezer_systems}
In this section, we illustrate the notion of coherent-classical estimation by a simple example involving the use  of a dynamic quantum squeezer as a coherent controller; e.g., see \cite{PET10Ca}. This example shows that the process of coherent-classical estimation has the potential to yield improved performance compared with purely classical estimation. 

An optical   cavity  consists of partially
reflecting mirrors arranged to produce a cavity mode when
coupled to a coherent light source; e.g., see \cite{BR04,GZ00}. By including a nonlinear optical element inside such a cavity, an  optical squeezer can be obtained. By using suitable linearizations and approximations, such an optical squeezer can be
 described by
the following quantum stochastic differential equations:
\begin{eqnarray}
\label{single_cavity}
da &=& -\frac{\gamma}{2}a dt -\chi a^* dt -  \sqrt{\kappa_1} d\mathcal{A}_1-  \sqrt{\kappa_2} d\mathcal{A}_3;
  \nonumber \\
d\mathcal{A}_1^{out} &=& \sqrt{\kappa_1}a dt + d\mathcal{A}_1; \nonumber \\
d\mathcal{A}_2^{out} &=& \sqrt{\kappa_2}a dt + d\mathcal{A}_2;
\end{eqnarray}
where $\kappa_1 > 0$, $\kappa_2 > 0$, $\gamma = \kappa_1+\kappa_2$,  $\chi \in \mathbb{C}$, and $a$ is a single
annihilation operator associated with the cavity mode; e.g., see \cite{BR04,GZ00}. This leads to a linear quantum system 
of the form
(\ref{sys}) as follows:
\begin{eqnarray}
\label{example_qsde}
\left[\begin{array}{l} d a(t)\\d a(t)^*\end{array}\right] &=& 
\left[\begin{array}{ll}
-\frac{\gamma}{2} & -\chi \\ -\chi^* & -\frac{\gamma}{2}
\end{array}\right]
\left[\begin{array}{l}  a(t)\\ a(t)^*\end{array}\right]dt 
\nonumber \\
&&
 -  \sqrt{\kappa_1}
\left[\begin{array}{l} d\mathcal{A}_1(t)
\\ d\mathcal{A}_2(t)^{\#} \end{array}\right];  \nonumber \\
&&
 -  \sqrt{\kappa_2}
\left[\begin{array}{l} d\mathcal{A}_2(t)
\\ d\mathcal{A}_2(t)^{\#} \end{array}\right];  \nonumber \\
\left[\begin{array}{l} d\mathcal{A}_1^{out}(t)
\\ d\mathcal{A}_1^{out}(t)^{\#} \end{array}\right] &=& 
\sqrt{\kappa_1}\left[\begin{array}{l}  a(t)\\ a(t)^*\end{array}\right]dt +
\left[\begin{array}{l} d\mathcal{A}_1(t)
\\ d\mathcal{A}_1(t)^{\#} \end{array}\right];\nonumber \\
\left[\begin{array}{l} d\mathcal{A}_2^{out}(t)
\\ d\mathcal{A}_2^{out}(t)^{\#} \end{array}\right] &=& 
\sqrt{\kappa_2}\left[\begin{array}{l}  a(t)\\ a(t)^*\end{array}\right]dt +
\left[\begin{array}{l} d\mathcal{A}_2(t)
\\ d\mathcal{A}_2(t)^{\#} \end{array}\right].\nonumber \\
\end{eqnarray}
Also associated with the squeezer system is the position operator $q = a+a^*$ and the momentum operator $p = \frac{a-a^*}{\imath}$.

\begin{figure}[htbp]
 \begin{center}
\includegraphics[width=6cm]{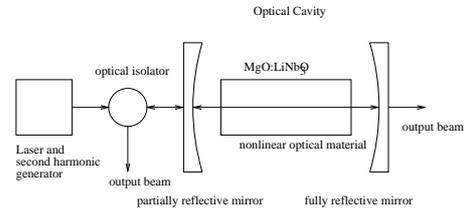}
 \end{center}
 \caption{An optical squeezer system.}
 \label{F1}
\end{figure}

\begin{figure}[htbp]
 \begin{center}
\psfrag{a}{$a$}
\psfrag{A1}{$\mathcal{A}_1$}
\psfrag{Ao1}{$\mathcal{A}_1^{out}$}
\psfrag{A2}{$\mathcal{A}_2$}
\psfrag{Ao2}{$\mathcal{A}_2^{out}$}
\includegraphics[width=6cm]{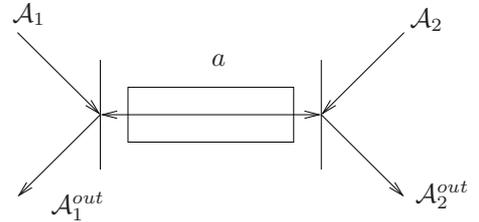}
 \end{center}
 \caption{Squeezer schematic diagram.}
 \label{F2}
\end{figure}

The construction of an optical squeezer is illustrated in Figure \ref{F1}. 
Such an optical squeezer is often represented as shown in the diagram in Figure \ref{F2}. 

\begin{figure}[htbp]
 \begin{center}
\psfrag{a}{$a$}
\psfrag{MA0}{$\mathcal{A}$}
\psfrag{MU}{$\mathcal{U}$}
\psfrag{MAT}{$\mathcal{\tilde A}$}
\psfrag{MY}{$\mathcal{Y}$}
\psfrag{yt}{$\tilde y$}
\psfrag{zh}{$\hat z$}
\includegraphics[width=8.5cm]{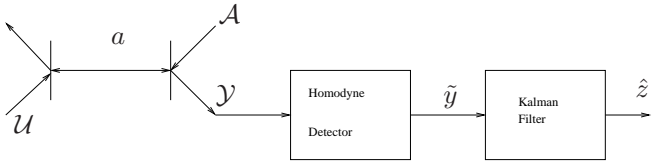}
 \end{center}
 \caption{Classical only estimation for an optical cavity plant.}
 \label{F2a}
\end{figure}

We now consider a quantum plant of the form (\ref{plant}) corresponding to a dynamic quantum squeezer. This system is described by the QSDEs
\begin{eqnarray}
\label{cavity_plant}
\left[\begin{array}{l} d a(t)\\d a(t)^*\end{array}\right] &=& 
\left[\begin{array}{ll}
-\frac{\gamma}{2} & -\chi \\ -\chi^* & -\frac{\gamma}{2}
\end{array}\right]
\left[\begin{array}{l}  a(t)\\ a(t)^*\end{array}\right]dt 
\nonumber \\&&
 -  \sqrt{\kappa_1}
\left[\begin{array}{l} d\mathcal{A}(t)
\\ d\mathcal{A}(t)^{\#} \end{array}\right];  \nonumber \\
&&
 -  \sqrt{\kappa_2}
\left[\begin{array}{l} d\mathcal{U}(t)
\\ d\mathcal{U}(t)^{\#} \end{array}\right];  \nonumber \\
\left[\begin{array}{l} d\mathcal{Y}(t)
\\ d\mathcal{Y}(t)^{\#} \end{array}\right] &=& 
\sqrt{\kappa_1}\left[\begin{array}{l}  a(t)\\ a(t)^*\end{array}\right]dt +
\left[\begin{array}{l} d\mathcal{A}(t)
\\ d\mathcal{A}(t)^{\#} \end{array}\right];\nonumber \\
z &=& \left[\begin{array}{ll}\frac{1}{\sqrt{2}} & -\frac{1}{\sqrt{2}} \end{array}\right]. 
\end{eqnarray}
This choice of $z$ corresponds to a scaled version of the momentum operator. We consider the case of $\chi = 0$ and $\kappa_1 = 0.5$, $\kappa_2 = 0.5$, $\gamma=1$. This case corresponds to a standard optical cavity without any squeezing. The matrices corresponding to the system (\ref{plant}) are
\begin{eqnarray*}
F &=& \left[\begin{array}{ll}-0.5 & 0 \\ 0 & -0.5 \end{array}\right];\nonumber \\
G_1&=&\left[\begin{array}{ll}-0.7071 & 0 \\ 0 & -0.7071 \end{array}\right]; \nonumber \\
G_2&=&\left[\begin{array}{ll}-0.7071 & 0 \\ 0 & -0.7071 \end{array}\right];\nonumber \\
H&=&\left[\begin{array}{ll}0.7071 & 0 \\ 0 & 0.7071 \end{array}\right]; \nonumber \\
K&=&\left[\begin{array}{ll}1 & 0 \\ 0 & 1 \end{array}\right];~C=\left[\begin{array}{ll}0.7071 & -0.7071  \end{array}\right]. 
\end{eqnarray*}
We then calculate the optimal classical only state estimator for this system using the standard Kalman filter equations (e.g., see \cite{KS72}) corresponding to a homodyne detector angle of $\theta =135^\circ$; i.e., measuring the momentum quadrature of $\mathcal{\tilde Y}$. This case is illustrated in Figure \ref{F2a}. 

\begin{figure}[htbp]
 \begin{center}
\includegraphics[width=8.5cm]{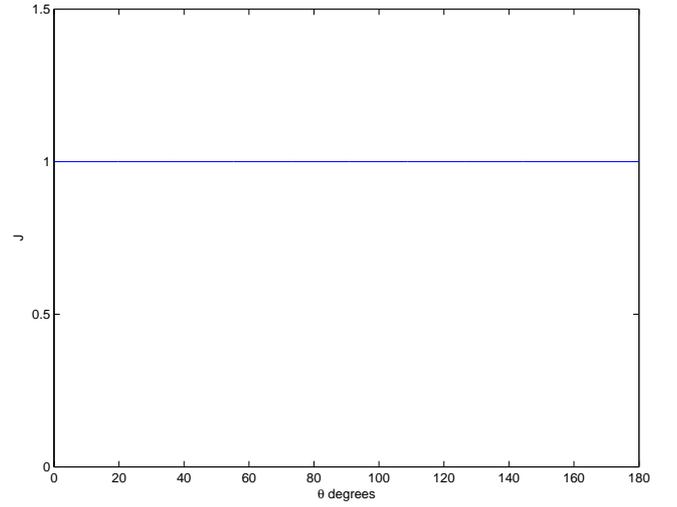}
 \end{center}
 \caption{Estimation error covariance versus homodyne detection angle $\theta$ for the case of classical only estimation.}
 \label{F3}
\end{figure}

The optimal classical only state estimator for this system leads to an error (\ref{cost}) of $J=1$. This is the same as the covariance of the variable $z$ without measurement. Also, the Kalman gain $G_e$ is found to be zero. That is, the measurement contains no information about the quantity $z$ to be estimated. This is consistent with Corollary 1 of \cite{PET13Ca} which notes that for a physically realizable annihilation operator system with only quantum noise inputs, any output field contains no information about the internal variables of the system. A similar result is found with any other homodyne detector angle $\theta$ as shown in Figure \ref{F3}. 

\begin{figure}[htbp]
 \begin{center}
\psfrag{a}{$a$}
\psfrag{ac}{$a_c$}
\psfrag{MA0}{$\mathcal{A}$}
\psfrag{MU}{$\mathcal{U}$}
\psfrag{MAT}{$\mathcal{\tilde A}$}
\psfrag{MY}{$\mathcal{Y}$}
\psfrag{MYT}{$\mathcal{\tilde Y}$}
\psfrag{yt}{$\tilde y$}
\psfrag{zh}{$\hat z$}
\includegraphics[width=8.5cm]{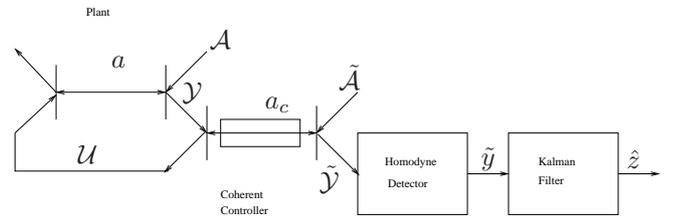}
 \end{center}
 \caption{Coherent-classical estimation for an optical cavity plant.}
 \label{F3a}
\end{figure}

\begin{figure}[htbp]
 \begin{center}
\includegraphics[width=8.5cm]{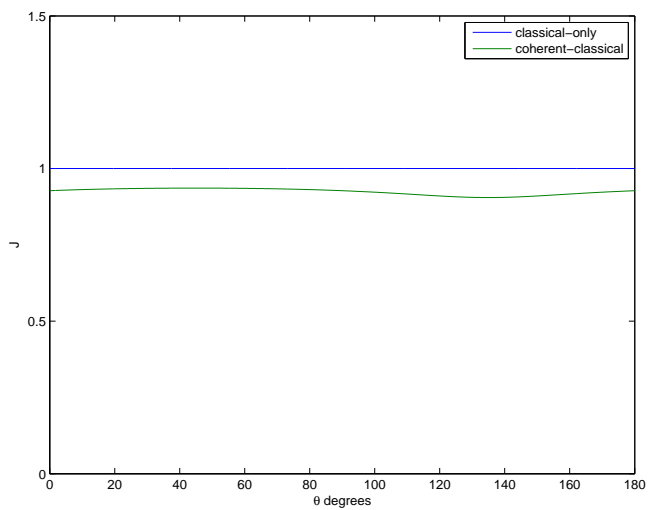}
 \end{center}
 \caption{Estimation error covariance versus homodyne detection angle $\theta$ for the case of coherent-classical estimation.}
 \label{F4}
\end{figure}

We now consider the case in which a dynamic squeezer is used as the coherent controller in  a coherent-classical estimation scheme. In this case, the coherent controller (\ref{coherent_estimator}) is described by the equations
\begin{eqnarray}
\label{squeezer_coherent_controller}
\left[\begin{array}{l} d a(t)\\d a(t)^*\end{array}\right] &=& 
\left[\begin{array}{ll}
-\frac{\gamma}{2} & -\chi \\ -\chi^* & -\frac{\gamma}{2}
\end{array}\right]
\left[\begin{array}{l}  a(t)\\ a(t)^*\end{array}\right]dt 
\nonumber \\&&
 -  \sqrt{\kappa_1}
\left[\begin{array}{l} d\mathcal{\tilde A}(t)
\\ d\mathcal{\tilde A}(t)^{\#} \end{array}\right];  \nonumber \\
&&
 -  \sqrt{\kappa_2}
\left[\begin{array}{l} d\mathcal{Y}(t)
\\ d\mathcal{Y}(t)^{\#} \end{array}\right];  \nonumber \\
\left[\begin{array}{l} d\mathcal{\tilde Y}(t)
\\ d\mathcal{\tilde Y}(t)^{\#} \end{array}\right] &=& 
\sqrt{\kappa_1}\left[\begin{array}{l}  a(t)\\ a(t)^*\end{array}\right]dt 
+
\left[\begin{array}{l} d\mathcal{\tilde A}(t)
\\ d\mathcal{\tilde A}(t)^{\#} \end{array}\right];\nonumber \\
\left[\begin{array}{l} d\mathcal{U}(t)
\\ d\mathcal{U}(t)^{\#} \end{array}\right] &=& 
\sqrt{\kappa_2}\left[\begin{array}{l}  a(t)\\ a(t)^*\end{array}\right]dt 
+
\left[\begin{array}{l} d\mathcal{Y}(t)
\\ d\mathcal{Y}(t)^{\#} \end{array}\right].\nonumber \\
\end{eqnarray}
Here, we choose $\chi = -0.5$ and $\kappa_1 = 5$, $\kappa_2 = 5$, $\gamma=10$. The matrices corresponding to the system (\ref{coherent_estimator}) are
\begin{eqnarray*}
F_c &=& \left[\begin{array}{ll}-5 & -0.5 \\ -0.5 & -5 \end{array}\right];\nonumber \\
G_{c1}&=&\left[\begin{array}{ll}-2.2361 & 0 \\ 0 &-2.2361 \end{array}\right]; \nonumber \\
G_{c2}&=&\left[\begin{array}{ll}-2.2361 & 0 \\ 0 &-2.2361 \end{array}\right];\nonumber \\
\tilde H_c&=&\left[\begin{array}{ll}2.2361 & 0 \\ 0 &2.2361 \end{array}\right]; \nonumber \\
\tilde K_{c1} &=&\left[\begin{array}{ll}1 & 0 \\ 0 & 1 \end{array}\right];~~
\tilde K_{c2}=0;\nonumber \\
H_{c} &=&\left[\begin{array}{ll}2.2361 & 0 \\ 0 &2.2361\end{array}\right];~~ 
K_{c1}=0;\nonumber \\
K_{c2}&=&\left[\begin{array}{ll}1 & 0 \\ 0 & 1 \end{array}\right].
\end{eqnarray*}
Then, the classical state estimator for this case is calculated according to equations (\ref{augmented_matrices}), (\ref{are1}), (\ref{classical_estimator_matrices}) for different values of the homodyne detector angle $\theta$. This case is illustrated in Figure \ref{F3a}. The resulting value of the cost $J$ in (\ref{cost}) along with the cost for the classical only estimator case is shown in Figure \ref{F4}. 

From this figure, we can see that the coherent-classical estimator performs  better than the classical only estimator with the best performance being achieved at a homodyne detector angle of $\theta = 135^\circ$ which corresponds to measuring the momentum quadrature of the field $\mathcal{\tilde Y}$. Note that a critical feature of this coherent-classical estimator is the use of coherent feedback. It does not seem possible to obtain improved performance with a coherent-classical estimator without the use of such feedback.  

\section{Conclusions}
\label{sec:conclusions}
In this paper, we have introduced the problem of classical-coherent estimation for quantum systems and shown via an example involving dynamic squeezers that the use of classical-coherent estimators can lead to significant improvement over the use of classical only estimators.


\end{document}